\documentclass[sigconf]{acmart}
\pdfoutput=1

\makeatletter                   
\def\mdseries@tt{m}             
\makeatother                    
\usepackage[plain]{fancyref}
\usepackage[draft=true]{minted} 
\usepackage{color}
\usepackage{hyperref}           
\hypersetup{
    colorlinks=true,
    linkcolor=blue,
    filecolor=red,      
    urlcolor=magenta,
    breaklinks=true,            
}
\usepackage{breakurl}           
\AtBeginDocument{%
  \providecommand\BibTeX{{%
    \normalfont B\kern-0.5em{\scshape i\kern-0.25em b}\kern-0.8em\TeX}}}

\setcopyright{none}
\makeatletter
\renewcommand\@formatdoi[1]{\ignorespaces}
\makeatother
\acmConference[KDD 2020 Workshop on Humanitarian Mapping]{KDD 2020 Workshop on Humanitarian Mapping}{}{}
\acmBooktitle{KDD 2020 Workshop on Humanitarian Mapping, San Diego, USA}
\acmPrice{none}
\acmISBN{}

\begin{document}
\sloppy  
\title{Mapping the South African health landscape in response to COVID-19}

\author{Nompumelelo Mtsweni}
\email{lelo@nompumelelo.me}
\affiliation{%
  \institution{Independent\\South Africa}
}

\author{Herkulaas MvE Combrink}
\orcid{0000-0001-7741-3418}
\affiliation{%
  \institution{University of the Free State\\
  Dept. Computer Science, University of Pretoria\\
  South Africa}
}
\email{CombrinkHM@ufs.ac.za}


\author{Vukosi Marivate}
\orcid{0000-0002-6731-6267}
\affiliation{%
  \institution{Dept. Computer Science, University of Pretoria\\
  Council for Scientific and Industrial Research\\
  South Africa}
}
\email{vukosi.marivate@cs.up.ac.za}

\renewcommand{\shortauthors}{Mtsweni and Combrink, et al.}

\begin{abstract}
When the COVID-19 disease pandemic infiltrated the world, there was an immediate need for accurate information. As with any outbreak, the outbreak follows a clear trajectory, and subsequently, the supporting information for that outbreak needs to address the needs associated with that stage of the outbreak. At first, there was a need to inform the public of the information related to the initial situation related to the “who” of the COVID-19 disease. However, as time continued, the “where”, “when” and “how to” related questions started to emerge in relation to the public healthcare system themselves. Questions surrounding the health facilities including COVID-19 hospital bed capacity, locations of designated COVID-19 facilities, and general information related to these facilities were not easily accessible to the general public. Furthermore, the available information was found to be outdated, fragmented across several platforms, and still had gaps in the data related to these facilities. To rectify this problem, a group of volunteers working on the covid19za project stepped in to assist. Each member leading a part of the project chose to focus on one of four problems related to the challenges associated with the Hospital information including: data quality, data completeness, data source validation and data visualisation capacity. As the project developed, so did the sophistication of the data, visualisation and core function of the project. The future prospects of this project relate to a Progressive Web Application that will avail this information for the public as well as healthcare workers through comprehensive mapping and data quality.
\end{abstract}



\keywords{COVID19,Public Health, Health Facilities, Mapping}

\begin{teaserfigure}
\centering
  \includegraphics[width=0.99\textwidth]{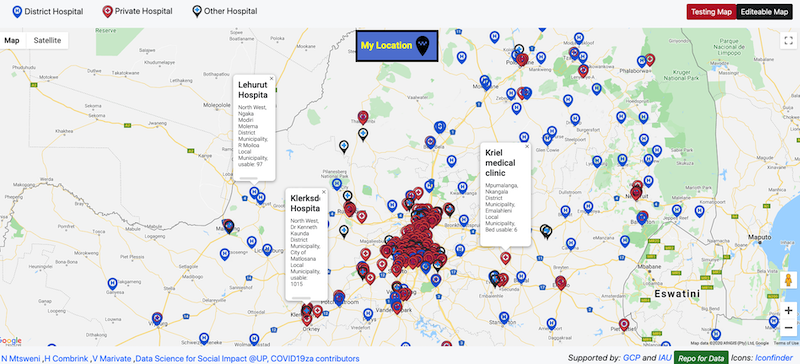}
  \caption{Proof of concept Health facility map of South Africa. \url{https://dsfsi.github.io/healthfacilitymap/}}
  \label{fig:teaser}
\end{teaserfigure}

\maketitle

\section{Introduction}
In the early phases of the COVID-19 disease in South Africa, with the number of infectious and infected people on the incline. During the early phases of the COVID-19 disease in South Africa, the Data Science for Social Impact (DSFSI) research group at the University of Pretoria and collaborators, experienced challenges related to information on a facility level related to the COVID-19 disease readiness \cite{marivate2020use}. 

As we move to the next phase of South Africa's experience of COVID-19 and its responses, we continue to look back at the progress of the Coronavirus COVID-19 (2019-nCoV) Data Repository for South Africa project~\cite{marivate_vukosi_2020_3819126}\footnote{\url{https://github.com/dsfsi/covid19za}}, from hereon referred to as the \emph{covid19za project}. The focus for this case study is on understanding the health system capacity in South Africa. A challenge that was identified early on in the project was the lack of updated health system data (data about hospitals, clinics and health resources). This challenge expanded when there was a realisation of the magnitude of work that would need to be done to collect and harmonise public health systems data for South Africa. This work builds on our initial work on collecting and collating COVID-19 data for South Africa \cite{marivate2020use}. In this case study, we discuss these challenges and present what has been achieved so far. 

\section{The beginning}

Initially, our goal was to collect information on the public healthcare system related to the South African COVID-19 response. We had anticipated that with the growth of cases, better health system information would be needed for more localised decision making. We collected the health system data for four primary reasons. We wanted to be able to identify: 
\begin{itemize}
    \item Where the COVID-19 hospitals were going to be in the country; 
    \item What the current situation in the country is related to the COVID-19 response;
    \item When the shortages were going to arise within the public healthcare system related to the COVID-19 response; 
    \item and how to assist the public in understanding which healthcare facilities were used for what specific response amidst the COVID-19 pandemic.
\end{itemize}

At the beginning of this task, these quetions where in their infance and fairly descriptive in nature. Initially, a few members in the \emph{covid19za project} volunteer group started asking questions about health system information (through posting issues on GitHub). These questions led to the intial directions of what would be tackled in building data towards mapping health systems information, capability and capacity in South Africa. We next describe our data collection and collation.

\section{Data}

\subsection{Sourcing the data}

The initial data contributed to the repository for health systems was a pivoted dataset, in a CSV format, containing the geo-coordinates of the locations of South African hospitals in the public sector and some of the hospitals in the private sector. This dataset is a subset of a much larger dataset from the South African National Department of Health (NDoH)’s Data Dictionary\footnote{\url{https://dd.dhmis.org/}}. The Data Dictionary provides a reference point for selected health information standards to support healthcare activities within South Africa. Unfortunately, there are differences in some of the variables in the data dictionary, lacking some of the vital information required for the COVID-19 response that have not been updated since 2018.

Although the data from the NDoH Data Dictionary listed a large number of facilities; it did not represent the current realities of some hospitals. There were some parts of the data that were outdated such as the hospital names(some hospital names have changed). The dataset also had a large number of missing data points including physical addresses and contact details. The facilities listed did not include attributes such as capacity (number of beds, number of Intensive Care Unit (ICU) beds), services available, and infrastructure (ambulances, ventilators, etc). Despite the fact that the dataset contains great details about the administrative division within which a facility resorts, it lacks information about health districts and clusters. More information about the health district approach in South Africa can be found at the Health Systems Trust~\cite{hst2001}\footnote{\url{https://www.hst.org.za/publications/}}. The Gauteng Province Cluster Policy defines the concept of health clusters \cite{gauteng2019} as \emph{'A collection of health facilities at different levels of healthcare provision (primary, secondary, tertiary and quatenary) which are grouped together in a designated demographic area.'} 

The Gauteng Cluster Policy did not only have details about health facilities in each cluster in Gauteng but also provided some details about the number of hospital beds in each cluster. It should be noted that some hospitals were duplicated (meaning one hospital could be in two different clusters) making the accounting of hospital beds per cluster questionable. This could lead to challenges when tallying capacity later.

Given that one of the \emph{covid19za project} contribtors, Anelda van der Walt, is involved in the Afrimapr\footnote{\url{https://afrimapr.github.io/afrimapr.website/}} the health systems task received further input and resources to use to collect further data. A prior study \cite{dell2016global} had looked at the geographical mal-distribution of surgical resources in South Africa. We reached out to the author and obtained the underlying data. The data is now available through a permissive creative commons license\cite{Dell2020} to allow and encourage re-use upon our request. The dataset has data about the number of beds (classified into which units of hospitals they belong to) in each hospital per province. This dataset has been the most useful in filling the gaps from the NDoH data dictionary. We next had to work on data cleaning and munging. 

\subsection{Data cleaning}

To maintain the integrity of the data, we had to work on maintaining the provenance of all of the data that is being collated in the dataset. Since the data is being collected from a wide variety of sources, and the data is available under a variety of licenses (or without licenses), we had to document data sources and ensure that we attribute people/organisations who have collected the data and are willing to share it. This route is chosen not just for attribution’s sake, but to encourage others to also open up their data to assist in such situations.

Combining various health facility lists to create a single, complete view of a country’s health system, is not a trivial task. For example, lists cannot be merged simply by using facility names. As mentioned before, facility names change over time and lists are often not updated. Furthermore, facilities may be listed under their full names, short names, and/or common names. Typos, capitalisation, misspelling, and varying naming conventions all contribute to the challenge of merging health facility lists by facility name. To understand some of the real-world challenges experienced in the analysis and merging of health facility lists and identify steps that can be streamlined for others wanting to do the same, we set out to assist with collating the South African data. Some aspects of the datasets and challenges experienced are discussed in work by Afrimapr\footnote{\url{http://www.afrimapr.org/blog/2020/merging-health-facility-lists-part1/}} and come with supporting R Markdown file\footnote{\url{https://github.com/anelda/merge_open_hospital_data/blob/master/reports/merge_open_hospital_data_part1.Rmd}}. The full report discussing these challenges is also available~\cite{anelda2020report}.

The work on the \emph{covid19za project} is also feeding into Afrimapr. Afrimapr is a Wellcome Open Research Fund project hosted at the Liverpool School of Tropical Medicine. The overarching aim of Afrimapr is to develop reusable open-source R building blocks to make open African health data more accessible. One of the focus areas at this stage is the afrihealthsites\cite{afrihealthsites2020}\footnote{\url{https://github.com/afrimapr/afrihealthsites}} package that currently provides easy access to two continental open health facility datasets - one developed by the KEMRI|Wellcome Trust Research Programme and hosted by WHO and a second collated by \url{healthsites.io}. Ongoing work includes adding functionality to include self-defined health facility lists (such as the ones described in this blog\footnote{\url{https://dsfsi.github.io/blog/mapping-healthsystem/}} post) for visualisation and analysis. 

\subsection{Gaps in the data}

We have made progress with getting data but it is still not enough. Currently,  there are still useful bits of information about hospitals that are still missing like clear hospital descriptions, their operational functions and most importantly the statuses on whether they are COVID-19 ready or not. Data about contact details of hospitals and contact personnel per hospital is also currently not complete. Another crucial missing bit of data in our healthcare facilities dataset is the estimate of population in some local municipalities and districts. Current bed numbers in hospitals are not verified as yet, the locations of the recently developed field hospitals are not available as yet and also a clear indication of which hospitals (both in the public and the private sector ) are being upgraded has not yet been made available.

\section{Inspiration from across the world}

\begin{figure}
\includegraphics[width=1\columnwidth]{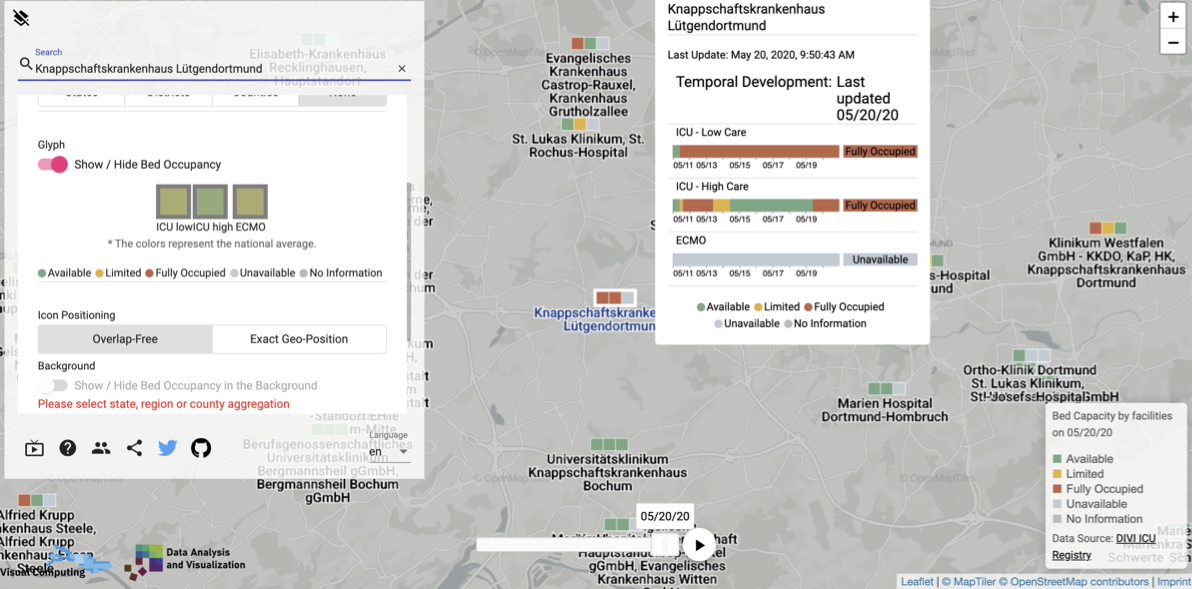}
\caption{German Health System Capacity Map}
\label{fig:german_map}
\end{figure}

During the COVID-19 pandemic, there have been many projects covering data analysis, modelling and mapping. Here we share a few mapping projects that have served as inspiration for some of the work that is being done on mapping the South African health system. One such project worked to map the availability of hospital resources in German hospitals. The project visualised all hospitals and provides easy to understand triage of the health system capacity for each hospital. An example is shown in Figure \ref{fig:german_map}

\begin{figure}
\includegraphics[width=\columnwidth]{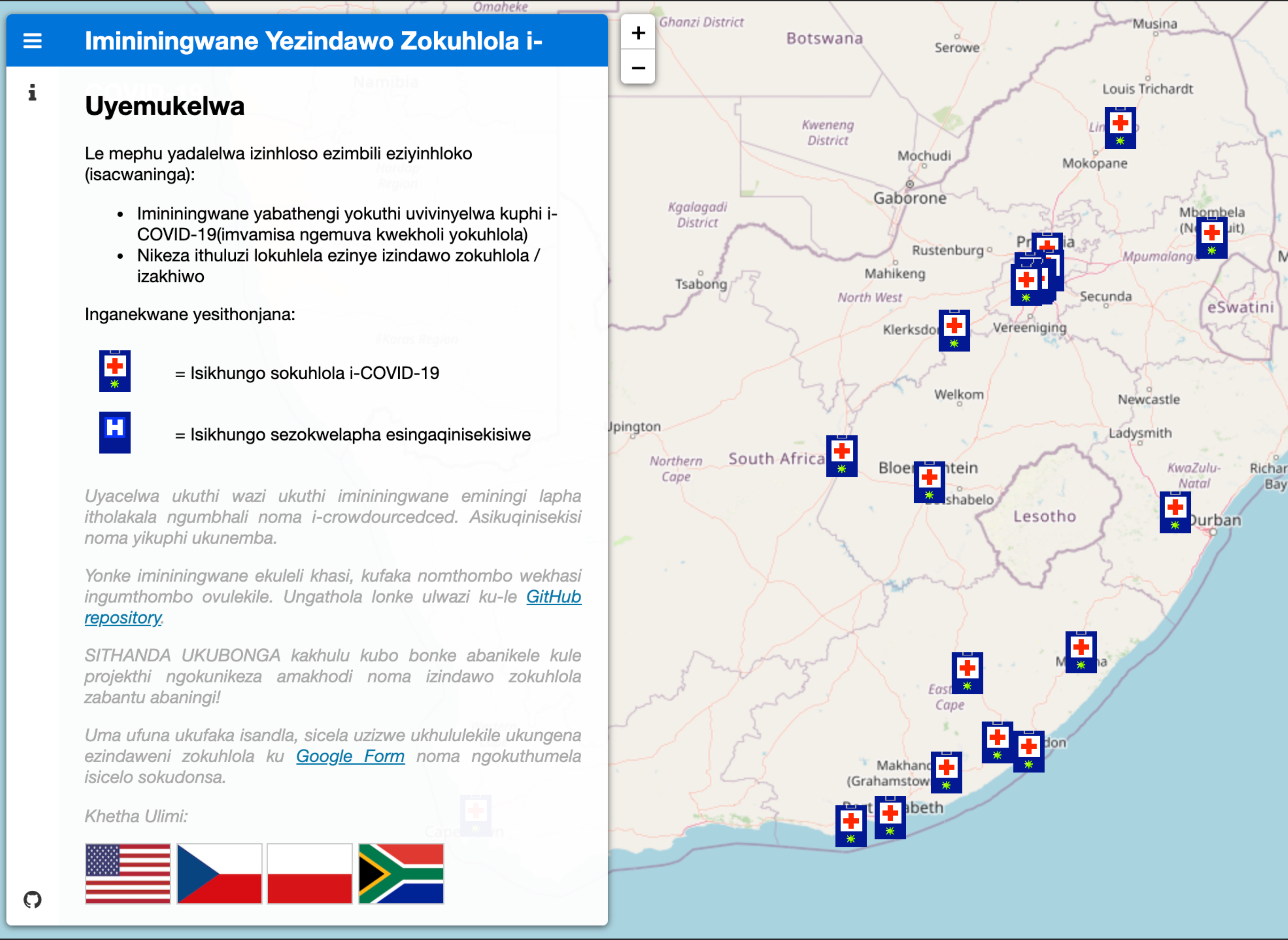}
\caption{Country testing map from \url{https://www.ineff.ch/cov19testmap/}}
\label{fig:testing_map}
\end{figure}

We were inspired by a COVID-19 testing facility map\footnote{\url{https://www.ineff.ch/cov19testmap/}} that visualised testing facilities in South Africa, Switzerland, Slovenia, Poland and the Czech Republic. One of the authors of this case study contributed to the translation of the map into isiZulu and also connected the map to the South African data in our project. The map is shown in Figure \ref{fig:testing_map}.

\section{Ongoing and future work}

Our ongoing work focuses on obtaining more sources of data to enhance the data that we already have. As the COVID-19 pandemic progresses in South Africa, we believe there will be more need for coordinating the health care system response. As such next discuss two sets of future work that we are planning. 

\subsection{Current state of mapping}

We created a mapping visualisation of the health systems data (as shown in \ref{fig:teaser}. Our next goal is to complete missing data and update the dataset with the current developments made by the South African governments. These developments include upgrading of public hospitals, additions of beds in some hospitals and development of field hospitals for triage. Populating the dataset with data related to these current developments will assist in preparing for the surge of  COVID-19 that is anticipated to peak in a few months~\cite{modelingZA2020}. 

\subsection{Tools for healthcare workers}

We are planning to develop a Progressive Web Application that will make accessibility of this data (amongst other data in the repository) easier. This will make it possible to get the information even to parts of the population that lives in remote  low-bandwidth areas where internet quality is not so strong or fast enough. This will be both for health care workers as well as the public. This will then open up the next phase of the project to develop tools for better understanding of individual health facility planning for the pandemic akin to the work on hospital projections \cite{zhang2020model}. 

\section{Conclusion}

The work described in this case study has been focused on how we have worked to collect, collate and clean health systems data for South Africa in response to the COVID-19 pandemic. This project emerged from the necessity to assess the healthcare system from a public point of view and inform not only the public, but the healthcare workers on the current situation within their facilities. Through this project, we have created our own maps of the health system (public and private) and also contributed to other mapping projects through making data available in an open manner. There is a large need for the NDoH to keep their health systems data up to date and easy to use and integrate with other systems both in public and private domains. This data becomes critical at a local management level if one needs to understand how well the health system is coping during a pandemic (we showed an example of the German ICU map that has a simple triage of each hospital). Furthermore, other innovations can be developed on top of the data (we showed an example of a testing facility map) that now includes South African (publicly) testing facilities that were announced by the Minister of Health, South Africa.

\begin{acks}
We would like to acknowledge Anelda van der Walt who has assisted in guding this work from the beginning and had numerous inputs into the process. We would like to acknowledge every person from the public that dedicated their time, effort and energy to assist during this pandemic. For the \emph{covid19za project} we would like to acknowledge all contributors. As this is an open contribution project, the updated list of contributors is available on the github repo\footnote{\url{https://github.com/dsfsi/covid19za/graphs/contributors}}. We would like to thank the DSFSI research group at the University of Pretoria for all their expertise, patience and hard work during this time. We also would like to thank all the employees of the NICD, DoH and WHO who assisted with data during this time.  We would like to acknowledge ABSA for sponsoring the industry chair and it's related activities to the project.
\end{acks}

\bibliographystyle{ACM-Reference-Format}
\bibliography{mapping-reference}

\end{document}